\documentclass[manuscript]{aastex}
\usepackage{emulateapj5}

\usepackage{epsfig}
\usepackage{amsmath}
\usepackage{amssymb}
\usepackage{natbib}
\usepackage{graphicx}

\shorttitle{Spin-up of millisecond pulsars}
\shortauthors{Bhattacharyya and Chakrabarty}

\begin{document}

\title{The Effect of Transient Accretion on the Spin-Up of Millisecond Pulsars}

\author{Sudip Bhattacharyya and Deepto Chakrabarty\altaffilmark{1}}
\affil{Department of Astronomy and Astrophysics, Tata Institute of
  Fundamental Research, 1 Homi Bhabha Road, Colaba, Mumbai 400005,
  India; sudip@tifr.res.in}

\altaffiltext{1}{Permanent address: MIT Kavli Institute for
  Astrophysics and Space Research, Massachusetts Institute of
  Technology, Cambridge, MA 02139, USA} 

\begin{abstract}
A millisecond pulsar is a neutron star that has been substantially
spun up by accretion from a binary companion. A previously
unrecognized factor governing the spin evolution of such pulsars is
the crucial effect of non-steady or transient accretion. We
numerically compute the evolution of accreting neutron stars through a
series of outburst and quiescent phases considering the drastic
variation of the accretion rate and the standard disk-magnetosphere
interaction.  We find that, for the same long-term average accretion
rate, X-ray transients can spin up pulsars to rates several times
higher than can persistent accretors, even when the spin down due to
electromagnetic radiation during quiescence is included.  We also
compute an analytical expression for the equilibrium spin frequency in
transients, by taking spin equilibrium to mean that no net angular
momentum is transferred to the neutron star in each outburst cycle. We
find that the equilibrium spin rate for transients, which depends on
the peak accretion rate during outbursts, can be much higher than that
for persistent sources.  This explains our numerical finding. This
finding implies that any meaningful study of neutron star spin and
magnetic field distributions requires the inclusion of the transient
accretion effect, since most accreting neutron star sources are
transients. Our finding also implies the existence of a submillisecond
pulsar population, which is not observed. This may point to the need
for a competing spin-down mechanism for the fastest-rotating accreting
pulsars, such as gravitational radiation.
\end{abstract}


\section{Introduction}\label{Introduction}

Millisecond pulsars (MSPs), a subset of fast-spinning neutron stars,
are an important probe of the physics of ultradense matter in compact
stellar cores
\citep{LattimerPrakash2007,Bhattacharyya2010,Bogdanovetal2007}.  When
radio MSPs were first discovered in the early 1980s, it was proposed
that they are spun up to high rates via accretion in low-mass X-ray
binaries \citep[LMXBs;][]{RadhakrishnanSrinivasan1982, Alparetal1982}.
This was eventually confirmed by discoveries of X-ray MSPs and
transitional pulsars \citep{WijnandsKlis1998, ChakrabartyMorgan1998,
  Archibaldetal2009, Papittoetal2013, deMartinoetal2013}.  However,
the detailed mechanism of this spin-up is not yet well understood.
One puzzling aspect is that the distribution of pulsar spin
frequencies cuts off sharply above around 730 Hz in both the X-ray
MSPs \citep{Chakrabartyetal2003, Chakrabarty2005, Chakrabarty2008,
  Patruno2010} and the radio MSPs \citep{FerrarioWickramasinghe2007,
  Hessels2008, Papittoetal2014}, well below the breakup spin rates for
  neutron stars \citep{Cooketal1994, Bhattacharyyaetal2016}. 
  Some authors have suggested the need for an additional
angular momentum sink such as gravitational radiation
\citep{Bildsten1998, Anderssonetal1999, Chakrabartyetal2003}. Others have argued that
standard magnetic disk accretion torque theory can account for the
spin distribution, for appropriate choices of pulsar magnetic field
strength $B$ and long-term average accretion rate $\dot M_{\rm av}$
\citep{Anderssonetal2005, LambYu2005, Patrunoetal2012b}. Detailed differences between
the radio and X-ray spin distributions
\citep{FerrarioWickramasinghe2007, Hessels2008, Papittoetal2014} led
to the suggestion of significant spin-down when the accretion phase
eventually ends and the binary ``detaches''
\citep{Tauris2012}. However, as we show, none of these analyses fully
accounted for the effect of transient accretion on the pulsar spin
evolution, even though most neutron star LMXBs are X-ray transients
\citep{Liuetal2013}, and among them almost all the X-ray MSPs are
transients \citep{Watts2012, PatrunoWatts2012}. 
As an example, while \citet{Possentietal1999} discussed the effects of 
transience on neutron star spin considering a lower $\dot M_{\rm av}$ for transients, 
one needs to consider the same values of parameters (including $\dot M_{\rm av}$)
for a persistent and a transient accretors in order to cleanly separate out the
effects of the transience phenomenon.

Many LMXBs alternate between long intervals of X-ray quiescence
lasting months to years, and brief transient outbursts lasting days to
weeks. These outbursts are believed to be caused by accretion
disk instabilities, and are seen in many systems.  Such instabilities
occur when $\dot M_{\rm av}$ is lower than a certain limit 
\citep[see, e.g.,][]{Lasota1997}.  When the steady mass
injection from a donor star accumulates enough mass in the disk, an
ionization instability is triggered in which the instantaneous
accretion rate $\dot M$ (and hence, the X-ray luminosity) increases by
several orders of magnitude, causing a transient outburst with low
duty cycle. When the accretion disk is emptied by the enhanced accretion rate
$\dot{M}$, the source returns to an extended X-ray quiescent state. A
new outburst occurs when sufficient mass accumulates in the disk again
\citep{Doneetal2007}.  The crucial effect of transient accretion on
the long-term spin-up of pulsars to millisecond periods has so far not
been reported.  In this paper, we show numerically  that, for a given
long-term average accretion rate $\dot{M}_{\rm av}$, the final $\nu$
value can be significantly different depending on whether the
accretion is persistent or transient.  We also analytically compute
the equilibrium spin frequency of neutron stars spun up in transient LMXBs.

\section{The Model}\label{Model}

\subsection{Disk-magnetosphere interaction and torques}\label{Torques}

Consider a spinning, magnetized neutron star accreting from a thin,
Keplerian disk.  The neutron star has gravitational mass $M$, radius
$R$, spin frequency $\nu$, and magnetic dipole moment $\mu = BR^3$
(where $B$ is the surface dipole magnetic field strength).  There are
three important length scales needed for understanding the different
accretion regimes.  The magnetospheric radius $r_m$, where the
magnetic and material stresses are equal, is
\begin{equation}\label{rm}
r_{\rm m} = \xi \left(\frac{\mu^4}{2 G M \dot M^2}\right)^{1/7} ,
\end{equation}
where $\xi$ is an order of unity constant that depends on details of the
disk-magnetosphere interaction \citep[see, e.g.,][]{Psaltis1999}.  The
corotation radius, where the stellar and Keplerian angular velocities
are equal, is
\begin{equation}
r_{\rm co} = \left(\frac{GM}{4\pi^2 \nu^2}\right)^{1/3} .
\end{equation}
Finally, the speed-of-light cylinder radius is $r_{\rm lc} =
c/2\pi\nu$.  For neutron stars, we may always assume $r_{\rm lc} >
r_{\rm co}$.  In the standard scenario for magnetic thin-disk
accretion, steady accretion occurs only when $r_{\rm m} < r_{\rm co}$ (the accretion phase),
with the magnetosphere lying within the corotation radius
\citep{PringleRees1972,Lamb1973}.  For $r_{\rm m} > r_{\rm co}$ (the
so-called ``propeller'' regime), accretion is largely shut off by a
centrifugal barrier \citep{Illarionov1975,Ustyugova2006}.  For $r_{\rm
  m} > r_{\rm lc}$, the accreted matter is swept clear of
magnetosphere, and the radio pulsar mechanism can turn on \citep{Stella1994}.
This has recently been confirmed with the discovery of three transitional
pulsars, which show radio pulsations in X-ray quiescent phases
\citep{Archibaldetal2009, Papittoetal2013, deMartinoetal2013}.
\citet{Linares2014} has listed the properties of their X-ray states, related to accretion,
as well as their pulsar state. These sources could be ideal to probe the 
accretion, propeller and quiescence phases.

The accretion/ejection and the interaction between the disk and 
the stellar magnetosphere exert a
torque on the neutron star that can be written as \citep[e.g.,][]{Parfreyetal2016}
\begin{equation}\label{Torque1}
	N = N_{\rm acc} + N_{\rm field},
\end{equation}
where $N_{\rm acc}$ is the contribution due to the accreting material, and
$N_{\rm field}$ is the contribution due to the interaction of the stellar magnetic
field with the disk. The specific angular momentum of the accreting matter at the radial distance $r = r_{\rm m}$
(i.e., disk inner edge) is $l = \sqrt{GMr_{\rm m}}$. Therefore, for $r_{\rm m} < r_{\rm co}$, 
$\dot M \sqrt{GMr_{\rm m}}$ is the rate of angular momentum added to the
neutron star, implying 
\begin{equation}\label{Torque2}
	N_{\rm acc} = \dot M \sqrt{GMr_{\rm m}}.
\end{equation}
For $r_{\rm m} > r_{\rm co}$, i.e., in the propeller regime, the accreted matter 
is expected to be largely thrown away from the system by the rotating stellar
magnetic field. This ejected matter takes away 
angular momentum from the neutron star, which can be of the order of $l$ for unit mass
of the accretion disk \citep{Tauris2012}. Here we note that, for $r_{\rm m}$ close to $r_{\rm co}$,
all accreting matter may not be expelled, and a portion of this gas may return to the
disk \citep[e.g.,][]{DAngeloetal2010}. However, as $\dot M$ varies rapidly during an
outburst of a transient source, $r_{\rm m}$ is expected to be considerably larger than 
$r_{\rm co}$ most of the time during the propeller phase. Considering the above points,
we can write
\begin{equation}\label{Torque3}
        N_{\rm acc} = -\eta \dot M \sqrt{GMr_{\rm m}},
\end{equation}
for $r_{\rm m} > r_{\rm co}$, where $\eta$, which is an order of unity positive constant,
includes the uncertainty due to unknown fraction of matter ejected for each $\dot M$ value.
In this paper, we consider a large range of $\eta$ ($0.2-1$), and show that 
our qualitative results and general conclusions do not depend on this value.

The torque on the neutron star due to the interaction of the stellar magnetic field
with the entire disk is \citep{Rappaportetal2004}
\begin{equation}\label{Torque4}
	N_{\rm field} = \int_{r_{\rm m}}^{\infty} B_z(r) B_{\phi}(r) r^2 {\rm d}r.
\end{equation}
Here, $B_z = \mu/r^3$, and $B_{\phi}$, the azimuthal component of the magnetic field,
appears due to the dragging of the magnetic field in the disk. Following \citet{Rappaportetal2004}
\citep[see also][]{LivioPringle1992, Wang1995}, we assume
$B_{\phi} = B_z (1 - \Omega/\Omega_{\rm K})$ for $r_{\rm m} \le r \le r_{\rm co}$,
and $B_{\phi} = -B_z (1 - \Omega_{\rm K}/\Omega)$ for $r \ge r_{\rm m} \ge r_{\rm co}$.
Here, $\Omega = 2 \pi \nu$ and $\Omega_{\rm K}$ is the Keplerian angular frequency.

Therefore, in the accretion phase,
\begin{equation}\label{Torque5}
\begin{split}
N_{\rm field} = \int_{r_{\rm m}}^{r_{\rm co}} B_z^2(r) [1 - \Omega/\Omega_{\rm K}(r)] r^2 {\rm d}r \\
- \int_{r_{\rm co}}^{\infty} B_z^2(r) [1 - \Omega_{\rm K}(r)/\Omega] r^2 {\rm d}r \\
	= \frac{\mu^2}{3 r_{\rm co}^3}\left[\frac{2}{3}-2\left(\frac{r_{\rm co}}{r_{\rm m}}\right)^{3/2}+\left(\frac{r_{\rm co}}{r_{\rm m}}\right)^3\right] \\
	= \frac{\mu^2}{9 r_{\rm m}^3}\left[2\left(\frac{r_{\rm m}}{r_{\rm co}}\right)^3-6\left(\frac{r_{\rm m}}{r_{\rm co}}\right)^{3/2}+3\right].
\end{split}
\end{equation}
This expression is same as the second term on the right hand side of the Equation (24) of \citet{Rappaportetal2004}.
In the propeller phase,
\begin{equation}\label{Torque6}
\begin{split}
	N_{\rm field} = -\int_{r_{\rm m}}^{\infty} B_z^2(r) [1 - \Omega_{\rm K}(r)/\Omega] r^2 {\rm d}r \\
	= -\frac{\mu^2}{9 r_{\rm m}^3}\left[3-2\left(\frac{r_{\rm co}}{r_{\rm m}}\right)^{3/2}\right].
\end{split}
\end{equation}
Note that, for $r_{\rm m} = r_{\rm co}$, this expression reduces to $-\mu^2/9 r_{\rm co}^3$,
that is the second term on the right hand side of the Equation (23) of \citet{Rappaportetal2004}.
Here is the reason why we use a more general expression.
\citet{Rappaportetal2004} considered that the disk always extends up to $r = r_{\rm co}$
in the propeller phase. This could be possible if $\dot M$ does not considerably vary.
But, $\dot M$, and hence $r_{\rm m}$, varies rapidly during an outburst of a transient source, which we consider.
In such cases, the disk is expected to either advance (during outburst rise) or recede
(during outburst decay) fast, and hence we consider a disk extending up to $r_{\rm m}$ in both accretion
and propeller phases \citep[e.g., ][]{Tauris2012}. Accordingly, the Equation~\ref{Torque6}
gives an appropriate torque by such an advancing and receding disk.

Therefore, in our computations, we use the following expressions of torque on the neutron star:
\begin{equation}\label{Torque7}
	N = \dot M \sqrt{GMr_{\rm m}} + \frac{\mu^2}{9 r_{\rm m}^3}\left[2\left(\frac{r_{\rm m}}{r_{\rm co}}\right)^3-6\left(\frac{r_{\rm m}}{r_{\rm co}}\right)^{3/2}+3\right]
\end{equation}
	for the accretion phase, and
\begin{equation}\label{Torque8}
	N = -\eta \dot M \sqrt{GMr_{\rm m}} -\frac{\mu^2}{9 r_{\rm m}^3}\left[3-2\left(\frac{r_{\rm co}}{r_{\rm m}}\right)^{3/2}\right]
\end{equation}
for the propeller phase.
Note that the accretion torque is positive for $r_{\rm m} < r_{\rm co}$ and negative for
$r_{\rm m} > r_{\rm co}$. Accretion thus drives the neutron star
toward an equilibrium where $r_{\rm m} = r_{\rm co}$ and the equilibrium
spin frequency is
\begin{equation}\label{equilibrium}
\nu_{\rm eq} = \frac{1}{2\pi}\sqrt\frac{GM}{r_{\rm m}^3} = 
\frac{1}{2^{11/14}\pi\xi^{3/2}}\left(\frac{G^5 M^5 \dot
      M^3}{\mu^6}\right)^{1/7} .
\end{equation}

In some of our numerical runs, we also consider other additional
spin-down mechanisms.  When $r_{\rm m} \simeq r_{\rm lc}$, some of our
runs include the electromagnetic (EM) torque due to magnetic dipole
radiation from the spinning neutron star, 
\begin{equation}\label{EMTorque}
N_{\rm EM} = - \frac{2\mu^2}{3r_{\rm lc}^3} = - \frac{16\pi^3\mu^2\nu^3}{3c^3}
\end{equation}
In addition, for all accretion regimes, some of the runs also include
the gravitational wave (GW) torque due to a rotating misaligned mass quadrupole
moment $Q$ \citep{Bildsten1998},
\begin{equation}\label{GWTorque}
N_{\rm GW} = - \frac{32GQ^2}{5}\left(\frac{2\pi\nu}{c}\right)^5 .
\end{equation}
We assume that $Q=0.5\times 10^{37}$ g~cm$^2$, consistent with the
upper limit set in the 401~Hz X-ray pulsar SAX J1808.4-3658
\citep{Hartmanetal2008}. 

\subsection{Transient outbursts}\label{Transient}

\begin{figure}[h]
\centering
\includegraphics*[width=12.0cm]{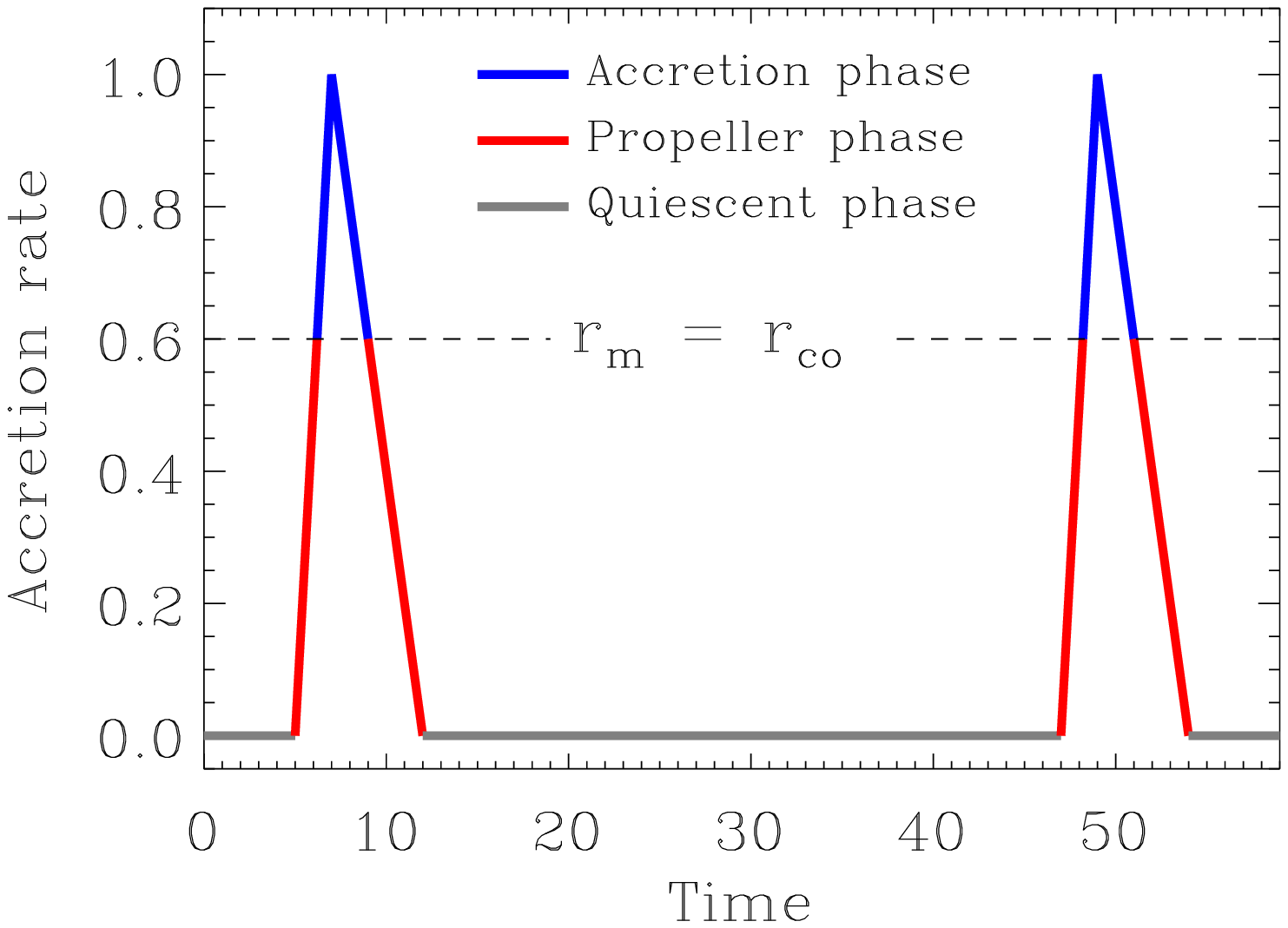}
\caption{A schematic illustration of how the accretion rate $\dot M$
  evolves through various phases for a transient LMXB. The instantaneous
  $\dot M$ is normalized by the outburst peak value
  $\dot{M}_{\rm max}$, and time is plotted in arbitrary units. Two
  outbursts are shown, with triangular outburst profiles, separated by
  a quiescent interval. The
  dashed horizontal line shows the condition $r_{\rm m} = r_{\rm co}$,
  which corresponds to the normalized effective accretion rate
  $\dot{M}_{\rm eff}/\dot{M}_{\rm max}$. When $\dot M$ is above this
  line, the source is in the accretion phase (blue). When $\dot M$ is below
  this line but within an outburst, the source is in the propeller
  phase (red). Outside of the two outbursts shown in the figure, the
  source is in the quiescent phase (grey).
\label{fig1}}
\end{figure}

For simplicity, we model the time evolution of a transient outburst as
a linear increase of $\dot M$ from quiescence ($\dot M\simeq 0$) to a
maximum value $\dot M_{\rm max}$, followed by a linear decrease of
$\dot M$ back down to quiescence (Figure~\ref{fig1}). In actuality, the
magnitude of the rise and decay slopes can be different, although this
does not affect our qualitative results. The evolution of $\dot M$ over an
outburst causes $r_{\rm m}$ to change as well. As the outburst rises, the
system moves from quiescence ($r_{\rm m} \simeq r_{\rm lc}$) through the
propeller regime ($r_{\rm m} > r_{\rm co}$) into the accretion regime
($r_{\rm m} < r_{\rm co}$).  During the decay, the systems passes back from
accretion through propeller into quiescence.  For an outburst duty
cycle (fractional duration) $f$, we have
\begin{equation}\label{transience}
  \dot M_{\rm av} \simeq \frac{1}{2}\,f\,\dot{M}_{\rm max} ,
\end{equation}
independent of the recurrence time.  It is convenient to define a
transience parameter $\dot m = \dot M_{\rm max}/\dot M_{\rm av}$; this
scales as $1/f$, with the proportionality factor depending on the
shape of the outburst light curve. This factor is $1/2$ for the triangular
outburst profiles we consider here.

\subsection{Numerical computation of spin evolution}\label{Numerical}

We wish to compare the spin evolution of persistent and transient
accretors for the same average accretion rate $\dot M_{\rm av}$. In
all cases, we start with a slowly spinning ($\nu\simeq 1$~Hz) neutron
star with mass $M=1.35 M_\odot$ \citep{ThorsettChakrabarty1999}, moment of inertia 
$I=0.4 MR^2$ \citep{RevnivtsevMereghetti2015}, and a
fixed surface magnetic field $B$.  We then compute the spin evolution of
the neutron star for a fixed $\dot M_{\rm av}$, using
Equations~\ref{Torque7} and \ref{Torque8}, and continue until a certain total rest mass
$\Delta M_{\rm tot}=0.6\,M_\odot$ is transferred to the neutron star.
Note that the amount of mass transferred to a neutron star depends on the
progenitor system \citep[e.g., LMXB versus intermediate-mass X-ray binary 
(IMXB);][]{Linetal2011,Chenetal2016}, and is not fully understood yet.
While a transferred mass of $\sim 0.1 M_\odot$ can make a neutron star
fast-spinning \citep[this paper; ][]{Taurisetal2013}, a gravitational mass as high as
$\sim 0.4 M_\odot$ could also be transferred \citep{Linetal2011}. Therefore,
since the transferred rest mass is higher than the corresponding gravitational mass
considering the binding energy \citep{Bagchi2011}, we continue our calculation
until $0.6\,M_\odot$ rest mass is transferred, to be on the safe side.
For the persistent case, we simply set the instantaneous accretion
rate $\dot M=\dot M_{\rm av}$.  For the transient case, we allow $\dot
M$ to vary through a series of outbursts with transience parameter
$\dot m$. The torques in
Equations~\ref{Torque7} and \ref{Torque8}, along with the instantaneous $\dot M$,
determine the amount of angular momentum $\Delta J$ and rest mass
transferred in each time step. We account for the conversion
of accreted rest mass to gravitational mass in the neutron star using
Equations 19 and 20 of \citet{Cipollettaetal2015}. Using the increased
$M$, we update our values of $R$ \cite[$\propto M^{-1/3}$; see,
  e.g.,][]{Ghosh2007} and $I$, and then proceed to the next time
step. Note that our choices for this relations are not unique, and in
general will depend upon the equation of state model assumed for the
neutron star core.  However, our qualitative results do not depend
on our specific choices here. 

We consider $\dot M_{\rm av}$ values ranging from $5\times 10^{14}$
to $3\times 10^{17}$ g~s$^{-1}$, corresponding to $\sim 0.0004-0.25\,\dot
M_{\rm Edd}$, where $\dot M_{\rm Edd}\simeq 1.2\times10^{18}$ g~s$^{-1}$ is the
Eddington critical accretion rate for a 1.35 $M_\odot$ neutron star.
We also consider transience parameters $\dot m$ in the 2--200 range.
These ranges are realistic. For example, the estimated $\dot M_{\rm av}$
and $\dot m$ are $\approx 7\times 10^{14}$ g~s$^{-1}$ and $\approx 40$ respectively
for SAX J1808.4--3658, and are $\approx 3\times 10^{16}$ g~s$^{-1}$ and $\approx 8$
respectively for 4U 1608--522 \citep[assuming a $0.2$ efficiency of energy generation;][]
{Wattsetal2008,Chakrabartyetal2003}.
Besides, we consider $B$ values in the range $3\times10^7 - 10^9$ G.
Note that $B$ is held fixed for any given run; we do
not model accretion-induced field decay, but rather start with a
fixed field strength that is already low (``post-decay'').  We
perform our runs both including and excluding the electromagnetic
torque term in Equation~\ref{EMTorque} during quiescence.

\section{Results}\label{Results}

Our results for a typical numerical run are shown in Figure~\ref{fig2}.
For the persistent accretor case, the spin frequency rapidly reaches
$\nu_{\rm eq}$ (Equation~\ref{equilibrium}) and then tracks $\nu_{\rm eq}$
as it gradually evolves with increasing $M$. For clarity, we will call
the persistent equilibrium spin $\nu_{\rm eq, per}$. We find that,
even for the same $\dot M_{\rm av}$, the evolution for a transient
accretor is very different. Since the instantaneous $\dot M$ varies
between a low value and $\dot M_{\rm max}$ over each transient
outburst, the instantaneous $\nu_{\rm eq}$ will also vary between a
low value and $\nu_{\rm eq, max}$ ($\nu_{\rm eq}$ corresponding to
$\dot M_{\rm max}$).  The $\nu_{\rm eq, max}$ curve is shown near the
top of the figure. The large swings in instantaneous $\nu_{\rm eq}$
occur on too short a time scale to plot in the figure. Moreover, the
outbursts are each much too short to allow the spin $\nu$ to track
these rapid swings in $\nu_{\rm eq}$. Instead, $\nu$ smoothly
increases until it reaches, and then tracks, an {\em effective}
equilibrium frequency $\nu_{\rm eq, eff}$ which is significantly
larger than $\nu_{\rm eq, per}$ (but not quite as large as $\nu_{\rm
  eq, max}$).  If we include an electromagnetic spin-down torque
during quiescence, then the resulting $\nu_{\rm eq, eff}^{\rm EM}$
curve is somewhat below the $\nu_{\rm eq,eff}$ curve (and with a
shallower slope), but still much above the curve for the persistent
case.  

Qualitatively, these results are quite general across all of our runs.
Figure~3 shows examples for a range of $B$, $\dot{M}_{\rm av}$, and
$\dot m = \dot{M}_{\rm max}/\dot{M}_{\rm av}$ to demonstrate the
robustness of our results.  Figures~\ref{fig3}a and \ref{fig3}b, which
are for $B = 10^8$ G and $\dot{M}_{\rm av} = 6.3\times10^{15}$
g~s$^{-1}$, show that $\nu_{\rm eq, max}$, $\nu_{\rm eq, eff}$ and
$\nu_{\rm eq, eff}^{\rm EM}$ are lower for lower $\dot m$. The effect
of EM torques is larger for higher $\dot m$, as seen in these figures,
for two reasons: first, EM torque has a strong spin dependence ($\sim
\nu^3$; Equation~\ref{EMTorque}), and second, the fractional
quiescence duration (during which the EM torque is active) decreases
with $f$, and hence increases with $\dot m$ (see
Equation~\ref{transience}).  Figure~\ref{fig3}c shows that all the 
spin frequencies are lower for a higher $B = 10^9$ G. This is because the
accretion disk does not penetrate as far into the magnetosphere, and
hence the minimum possible $r_{\rm m}$ value is higher
(Equation~\ref{rm}). However, $\nu_{\rm eq, eff}$ is still
significantly larger than $\nu_{\rm eq, per}$ for this case. In
Figure~\ref{fig3}d, we consider a much lower $B = 3\times10^7$ G and a
smaller $\dot{M}_{\rm av}$, and find that all our results from
Figure~\ref{fig2} are still valid. Here, we also show that the
gravitational wave torque (Equation~\ref{GWTorque}) can bring down
$\nu$ significantly (in this case, below the observed cut-off value of
$\simeq 730$Hz), but $\nu$ is still much higher than $\nu_{\rm eq,
  per}$.

In order to examine the effect of the uncertainty in $\eta$ value (Equation~\ref{Torque8}),
we compute spin evolution for $\eta = 0.2, 0.5$ and $1$,
keeping other parameter values same (see Figure~\ref{fig4}). This figure
shows that not only the nature of spin evolution remains same for this
wide range of $\eta$ values, but also quantitatively the curves are not 
very different from each other. For example, there is $\approx 12$\%
difference in $\nu$ values between $\eta = 0.2$ and  $\eta = 1$, after $0.6 M_\odot$
rest mass is added to the neutron star. 
Therefore, our general conclusion, that $\nu$ attains a higher value
for transience for the same $\dot M_{\rm av}$, remains valid
for the assumed large range of $\eta$ values.
Moreover, for a lower value of $\eta$, the spin-down torque is lower
(Equation~\ref{Torque8}), and hence the star acquires an even higher $\nu$.

Figure~\ref{fig5} confirms the above conclusions in a compact manner and
for a wide range of $\dot M_{\rm av}$ and $\dot m$ values. This figure
comprehensively shows that the more extreme the transient behavior
is (larger $\dot m$), the faster a neutron star spins for a given
$\Delta M_{\rm tot} (= 0.6 M_\odot)$. This figure also shows that the effect of EM
torques can be significant if $\nu$ or $\dot{m}$ has a high value.

\section{Analytical calculation of equilibrium spin frequency for
  transients}\label{Analytical}  

Figures~\ref{fig2} and \ref{fig3} show that a transient source attains the
equilibrium spin frequency $\nu_{\rm eq, eff}$, which is several times
higher than that ($\nu_{\rm eq}$) for a persistent source. 
It is also interesting to note that $\nu_{\rm eq, eff}$ and
$\nu_{\rm eq,max}$ maintain a constant ratio, as shown by the
horizontal part of the red dash-dot curve of Figure~\ref{fig6}b.
In this section, we will try to analytically understand these two new results.
Moreover, the analytical expression for $\nu_{\rm eq}$ for persistent accretors
(Equation~\ref{equilibrium}) is widely used to understand the spin
distribution of MSPs, even though most of the neutron star LMXBs are
transients. It would, therefore, be preferable to find an analytical expression of
$\nu_{\rm eq, eff}$ appropriate for transient accretors, which could
then be used to better understand the spin evolution and distribution
of MSPs, most of which evolved in transients. 

The spin-equilibrium condition for a transient source is somewhat different from that for a
persistent source.
Here is the reason. A persistent source, when it reaches the spin equilibrium, is expected to
always remain in the spin equilibrium, as the $r_{\rm m} = r_{\rm co}$ condition could
continuously remain valid. This spin equilibrium frequency $\nu_{\rm eq}$, given
in Equation~\ref{equilibrium}, evolves with increasing $M$ (Section~\ref{Results}).
On the other hand, $\dot{M}$, and hence $r_{\rm m}$, change drastically 
for a transient source during an outburst. As a result, the neutron star always gains
angular momentum during the accretion phase (Equation~\ref{Torque7}) and loses
angular momentum during the propeller phase (Equation~\ref{Torque8}).
Therefore, unlike in the case of a persistent source, a spin equilibrium
cannot not be established at every instant for a transient source, and
$r_{\rm m} = r_{\rm co}$ is not 
the correct condition for spin equilibrium of transients, as $r_{\rm m}$
drastically evolves. A simple balance of the positive torque (Equation~\ref{Torque7})
with the negative torque (Equation~\ref{Torque8}) also does not work, because
they cannot balance each other throughout an outburst, as $r_{\rm m}$ evolves.
What could then be the criterion for the spin equilibrium of a transient accretor? Note that,
although $\nu$ evolves throughout the two phases of a 
given outburst in a cyclic manner, this change is negligible, given that a typical outburst
duration is very small compared to the spin-up timescale of a neutron star. Therefore,
if no net angular momentum is added to the neutron star in an outburst
cycle, the small cyclic change in $\nu$ during each outburst can be ignored.
As a result, for time scales longer than an outburst duration, a spin equilibrium for
transients can be established if the stellar angular momentum gain in the
accretion phase cancels the stellar angular momentum loss in the propeller 
phase in every outburst cycle.
This criterion, to the best of our knowledge, has not previously been used to
calculate the spin equilibrium frequency for transients.

In order to analytically estimate the equilibrium spin frequency $\nu_{\rm eq,eff}$,
we consider a simple but general torque formula
\begin{equation}\label{Torque9} 
\frac{{\rm d}J}{{\rm d}t} = \pm A \dot{M}^n,
\end{equation}
where $J$ is the stellar angular momentum, $A$ is a positive constant, and the
positive and negative signs correspond to $r_{\rm m} < r_{\rm co}$ and
$r_{\rm m} > r_{\rm co}$ respectively. 
We use this torque formula, because torques given by 
Equations~\ref{Torque7} and \ref{Torque8} are not simple enough for analytical calculations.
However, before proceeding further, it is desirable to check if the form of the torque
formula given in Equation~\ref{Torque9} is reasonable, that is if it can be approximately constructed from
Equations~\ref{Torque7} and \ref{Torque8}. In order to do this, we note that 
$N_{\rm field}$ tends to $\mu^2/3r_{\rm m}^3$ for $r_{\rm m} << r_{\rm co}$ in the
accretion phase, and $N_{\rm field}$ tends to $-\mu^2/3r_{\rm m}^3$ for $r_{\rm m} >> r_{\rm co}$ 
in the propeller phase (Equations~\ref{Torque5} and \ref{Torque6}). 
For other values of $r_{\rm m}$, $N_{\rm field}$ has a 
value in between these limiting values. Therefore, using 
Equations~\ref{Torque5} and \ref{Torque6}, one could approximately write,
\begin{equation}\label{Torque10}
	N_{\rm field} = \pm \frac{\beta \mu^2}{3r_{\rm m}^3},
\end{equation}
where $0 \le \beta \le 1$, and the positive (negative) sign is for the accretion (propeller) phase.
Similarly, using Equations~\ref{Torque2} and \ref{Torque3}, and for $\eta = 1$, one can write,
\begin{equation}\label{Torque11}
	N_{\rm acc} = \pm \dot M \sqrt{GMr_{\rm m}},
\end{equation}
where the positive (negative) sign is for the accretion (propeller) phase.
Therefore, using Equation~\ref{Torque1}, as well as Equations~\ref{Torque10} and \ref{Torque11}, 
the approximate torque can be written as
\begin{equation}\label{Torque12}
	N = \pm \dot M \sqrt{GMr_{\rm m}} \pm \frac{\beta \mu^2}{3r_{\rm m}^3}.
\end{equation}
Using the Equation~\ref{rm}, it is easy to verify that this torque formula
is exactly of the form given in Equation~\ref{Torque9} (for $n = 6/7$).
Moreover, a comparison between the blue solid and red dash-dot curves of 
Figure~\ref{fig6}a shows that Equation~\ref{Torque12} gives a spin evolution
very similar to that given by Equations~\ref{Torque7} and \ref{Torque8},
with a quantitative difference at the level of a few percent. Note that,
following \citet{Tauris2012}, we assume $\beta = 1/3$ in Equation~\ref{Torque12}
for Figure~\ref{fig6}a. However, this specific value
of $\beta$ is only for demonstration, and the
spin evolution curve changes at most by a few percent between $\beta = 0$ and $\beta = 1$.
Therefore it is reasonable to proceed with the simple but general torque formula given by 
Equation~\ref{Torque9}, which is suitable to analytically estimate the equilibrium spin
frequency $\nu_{\rm eq,eff}$. Note that this exercise
will also be very useful to understand the results described in Section~\ref{Results}.

As mentioned earlier, the spin frequency attains the equilibrium value
$\nu_{\rm eq,eff}$, if $\Delta J$ in the accretion phase ($r_{\rm m} <
r_{\rm co}$) and $\Delta J$ in the propeller phase ($r_{\rm m} >
r_{\rm co}$) during an outburst cancel each other. Here, using Equation~\ref{Torque9},
\begin{equation}\label{Integral} 
\Delta J = \int {\rm d}J = \pm A \int \dot{M}^n {\rm d}t = \pm A_1
\int \dot{M}^n {\rm d}\dot{M}, 
\end{equation}
where ${\rm d}\dot{M}/{\rm d}t$, and hence $A_1 = A/({\rm
  d}\dot{M}/{\rm d}t)$, are constants for a linear $\dot{M}$ profile
during an outburst.  Therefore the above mentioned requirement for
$\nu = \nu_{\rm eq,eff}$ gives 
\begin{equation}
\dot M_{\rm max}^{n+1} - \dot M_{\rm
  eff}^{n+1} = \dot M_{\rm eff}^{n+1} - \dot M_{\rm lc}^{n+1} , 
\end{equation}
and hence
\begin{equation}\label{Eff1} 
\frac{\dot M_{\rm max}}{\dot M_{\rm eff}} = 2^{1/(n+1)},
\end{equation}
where the effective accretion rate $\dot M_{\rm eff}$ corresponds to
$r_{\rm m} = r_{\rm co}$, and we assume $\dot M_{\rm lc}$
corresponding to quiescence ($r_{\rm m} \simeq r_{\rm lc}$) is
zero. Since the equilibrium spin frequency scales as $\nu_{\rm eq}
\propto \dot M^{3/7}$ (Equation~\ref{equilibrium}), we find
\begin{equation}\label{Eff2} 
\frac{\nu_{\rm eq,eff}}{\nu_{\rm eq,max}} = 2^{-3/(7(n+1))}.
\end{equation}
Equation~\ref{Eff2} gives an analytical expression of equilibrium spin
frequency $\nu_{\rm eq,eff}$ for transient sources. This equation also
shows why $\nu_{\rm eq,eff}/\nu_{\rm eq,max}$ is a constant, as shown in
Figure~\ref{fig6}b. We expect $\nu_{\rm eq,eff}/\nu_{\rm eq,max}
\approx 0.85$ for the torque in Equation~\ref{Torque9}, with $n = 6/7$. 
This expected value from our simple analytical calculation very
well matches (within $4$\%) with the value from our numerical computation with 
the approximate torque given in Equation~\ref{Torque12}
(Figure~\ref{fig6}b). Even the constant $\nu_{\rm eq,eff}/\nu_{\rm eq,max}$ ratio
from the numerical computation with the torques given in Equations~\ref{Torque7}
and \ref{Torque8} matches within $10$\% with the analytical value of 0.85. Note that
this matching is better for a lower value of $\eta$ in Equation~\ref{Torque8}.
Therefore, our analytical results for a simple torque formula is valid
for the realistic torques (Equations~\ref{Torque7} and \ref{Torque8})
with a small error of a few percent.

Finally, we analytically estimate the range of $\nu_{\rm eq,eff}/\nu_{\rm eq,per}$, 
i.e., the ratio of spin rates
for transient and persistent accretors with the same $\dot{M}_{\rm av}$ value.
For this, we consider a typical $\dot{M}_{\rm max}/\dot{M}_{\rm av}$ of $10-100$ \citep{Burderietal1999}.
Since $\nu_{\rm eq} \propto \dot M^{3/7}$ (Equation~\ref{equilibrium}), this range
of $\dot{M}_{\rm max}/\dot{M}_{\rm av}$ gives a $\approx 2.7-7.2$ range for
$\nu_{\rm eq,max}/\nu_{\rm eq,per}$. Assuming $n = 6/7$ in Equation~\ref{Torque9},
and hence, $\nu_{\rm eq,eff}/\nu_{\rm eq,max} = 0.85$, we analytically estimate 
a $\sim 2-6$ range of $\nu_{\rm eq,eff}/\nu_{\rm eq,per}$. 
This is consistent with the numerically computed curves displayed in Figures~\ref{fig2} and \ref{fig3}.

\section{Discussion of assumptions}\label{Assumptions}

We use $\xi = 1$ in this paper, which is consistent with the expected range 
$\sim 0.5-1.4$ \citep{Wang1996}. Note that a different value of $\xi$ does not
change our qualitative results, as $\nu_{\rm eq,per}$, $\nu_{\rm eq,max}$, and hence
$\nu_{\rm eq,eff}$ scale with $\xi$ in the same way, i.e., $\propto \xi^{-3/2}$.
Note that, for fixed values of the equilibrium spin frequency and other parameters, 
the magnetic field $B \propto \xi^{-7/4}$ (Equation~\ref{equilibrium}). Consequently,
in order to attain a measured $\nu$ value, the stellar magnetic field is to be lower
for a higher value of $\xi$. Therefore, a value of $\xi$ different from $1$ will not
change our conclusions, but our assumed $B$ values will be different.

We use a linear $\dot{M}$ profile during both outburst rise and decay,
because this is the simplest and the cleanest profile for the
demonstration of our results. In reality, both rise and decay profiles
may be complex, can have several peaks, can have a somewhat flat top
and may be difficult to fit with a simple function \citep[see, e.g.,
  Figure 2 of][]{YanYu2015}. However, such a complex profile will not
in general change our conclusions. For example, an exponential decay
profile ($\dot{M} \propto \exp[{-t/\tau}]$ with time constant $\tau$)
gives $\nu_{\rm eq,eff}/\nu_{\rm eq,max} = 2^{-3/7n}$, which is 0.71
for $n = 6/7$.  Therefore, for a linear rise and an exponential
decay profile, which is often seen, the value of $\nu_{\rm
  eq,eff}/\nu_{\rm eq,max}$ should be between 0.71 and 0.85 for $n 
= 6/7$.  For a profile having a flat top, $\nu_{\rm eq,eff}/\nu_{\rm
  eq,max}$ is expected to have a higher value. 

Is it justified to keep $B$, $\dot M_{\rm av}$ and $\dot M_{\rm max}$
fixed in our calculation of evolution? We consider each parameter in
turn. In the case of $B$, it is convenient to keep the parameter fixed
in order to cleanly demonstrate the effect of transient accretion on
the spin evolution. We note that $B$ likely decreases by orders of
magnitude from an initial high value ($\sim 10^{12}$ G) on a time
scale short compared to the LMXB lifetime \citep[see,
e.g,][]{Pageetal2000, GeppertRheinhardt2002, LambYu2005, Patrunoetal2012a, IstominSemerikov2016}, and
hence the use of a fixed low post-decay $B$ value may be justified. We
also find this with our numerical calculations of spin evolution for
two initial $\nu$ values, 1 Hz and 100 Hz, keeping other parameter
values same.  By the time the star is spun up to 100 Hz, $B$ must
decrease to a much lower value or else the star could not be spun up
to this high $\nu$ value, and a further major decrease of the $B$
value is unlikely. Since we find that the spin evolution curves for
both cases are very similar to each other, we conclude that a fixed
$B$ value considered in numerical calculation does not have an impact
on the general conclusions of this paper.

In the case of $\dot M_{\rm av}$, we hold the parameter fixed in order
to separate out the effect of transient accretion from the much slower
variation of $\dot M_{\rm av}$ due to binary evolution, which is also
already a much better studied issue.  We do not expect any slow
evolution of $\dot 
M_{\rm av}$ to change the main findings of this paper. Finally, in
the case of $\dot M_{\rm max}$, we hold this parameter fixed merely
for the purpose of demonstration.  In reality, $\dot M_{\rm max}$
varies \citep[usually within a factor of 10;][]{YanYu2015}, and
depending on this, $\nu$ should track an average $\nu_{\rm eq,eff}$
(e.g., in between red dashed curves of Figures~\ref{fig3}a and
\ref{fig3}b).  However, our conclusions are not affected by this.

\section{Summary and implications}\label{Implications}

\begin{figure}[h]
\centering
\includegraphics*[width=10.0cm]{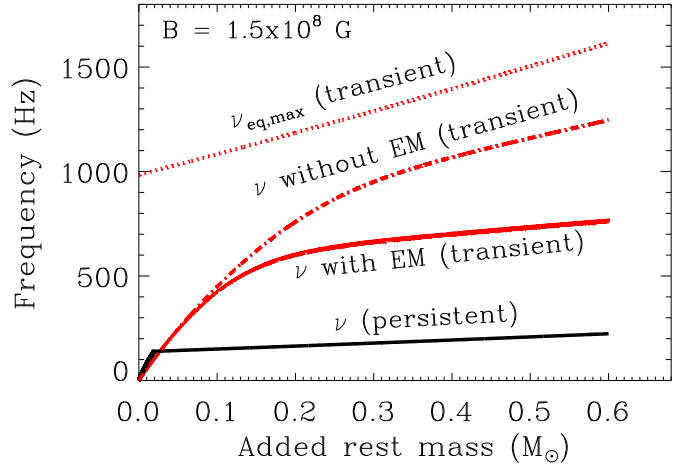}
\caption{
Numerically computed evolution of spin frequency
  versus transferred rest mass. We assume initial parameter
  values of gravitational mass $M = 1.35 M_\odot$ and $\nu = 1$~Hz, and fixed parameter
  values of $B = 1.5\times10^8$~G and $\dot{M}_{\rm av} =
  6.3\times10^{15}$ g~s$^{-1}$ (i.e., $10^{-10}$ 
  $M_\odot$/yr). The torques given in the Equations~\ref{Torque7} and \ref{Torque8} (with $\eta = 1$)
  are used. This figure compares the spin rate
  evolution between a transient and a persistent source for the same average accretion rate. 
  The three upper curves (in red) are for a transient with the transience
  parameter $\dot{m} = \dot M_{\rm max}/\dot M_{\rm av} = 100$.  Among
  these, the dotted curve corresponds to the maximum possible
  equilibrium spin frequency $\nu_{\rm eq, max}$ (which is $\nu_{\rm
    eq}$ for $\dot M_{\rm max}$), the dash-dot curve corresponds to the
  spin frequency $\nu$ without considering the effect of EM
  torques, and the solid curve corresponds to the spin
  frequency $\nu$ including  spin down due to EM torques. The
  lower curve (in black) is the spin frequency for a persistent
  accretor with $\dot{M} = \dot M_{\rm av}$. The nearly saturated value after
  the initial rise corresponds to the equilibrium spin frequency
  $\nu_{\rm eq, per}$ (which is $\nu_{\rm eq}$ for persistent
  accretion). Note that the persistent accretor needs only a 
  small amount of transferred mass to attain $\nu_{\rm eq, per}$.
  This figure shows that the neutron star in a transient
  can spin up to a much higher value relative to that in a persistent
  source.  
\label{fig2}}
\end{figure}

\begin{figure}[h]
\centering
\includegraphics*[width=6.0cm]{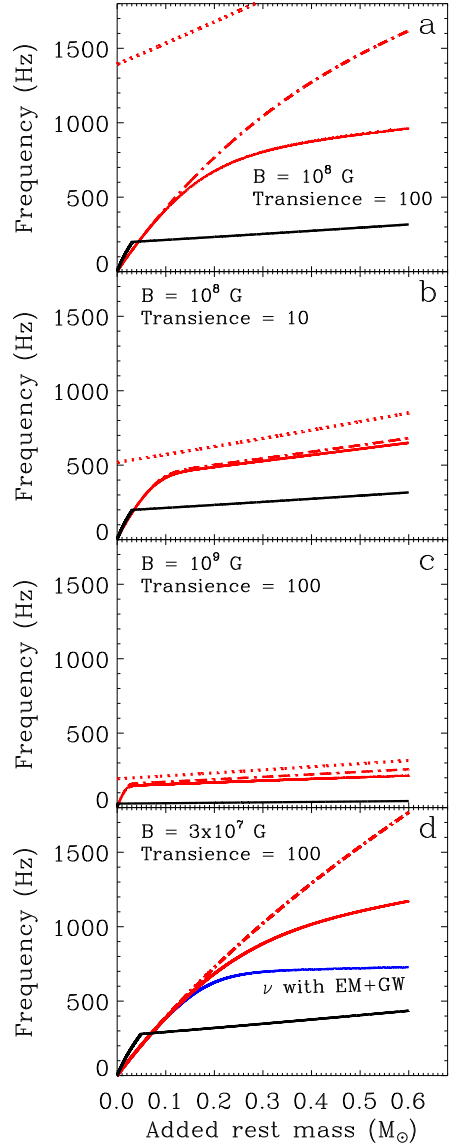}
\vspace*{-0.5in}
\caption{Numerically computed evolution of spin frequency as in
  Figure~\ref{fig2}, for several different cases. {\it Panel-a}: same
  as in Figure~\ref{fig2}, except with $B = 10^8$~G.  {\it Panel-b}:
  same as in {\it panel-a}, except with transience parameter $\dot{m}
  = 10$. {\it Panel-c}: same as in {\it panel-a}, except with $B =
  10^9$~G.  {\it Panel-d}: same as in {\it panel-a}, except with $B =
  3\times10^7$~G and $\dot M_{\rm av} = 1.2\times10^{15}$ g~s$^{-1}$
  (i.e., $\approx 10^{-3}\,\dot M_{\rm Edd}$ for a 1.35 $M_\odot$
  neutron star).  The additional blue solid curve in this panel
  includes the spin down due to both EM and GW. This figure confirms
  the findings from Figure~\ref{fig2} for a wide range of parameter
  values.
\label{fig3}}
\end{figure}

The main finding of this paper is that the spin rate of a transient
LMXB pulsar attains a much higher value than that for a persistent
LMXB with the same average accretion rate, usually even for less than $0.1
M_\odot$ mass transferred to the neutron star.  This is easily visible
in Figures~\ref{fig2} and \ref{fig3}. This crucial effect of transient
accretion on the spin-up of neutron stars implies that any meaningful
study of the observed spin distribution of MSPs requires its
inclusion.  This effect will also have impact on the current
understanding of spin-up and spin-down torques, accretion, binary evolution and
$B$-values, because nearly all the accreting MSPs are transients.

In this paper, we also report, for the first time, an analytical
expression of equilibrium spin frequency appropriate for transients.
The standard expression of $\nu_{\rm eq}$ in persistent accretors
(Equation~\ref{equilibrium}) laid the foundation for previous pulsar
spin distribution studies.  However, this should now be replaced by
our expression for $\nu_{\rm eq,eff}$ (Equation~\ref{Eff2}) for
transients. Note that $\nu_{\rm eq,eff}$, unlike $\nu_{\rm eq}$,
depends on the torque law, and hence may provide a way to better
understand the interaction between the accretion disk and the pulsar
magnetosphere.

\begin{figure}[h]
\centering
\includegraphics*[width=10.0cm]{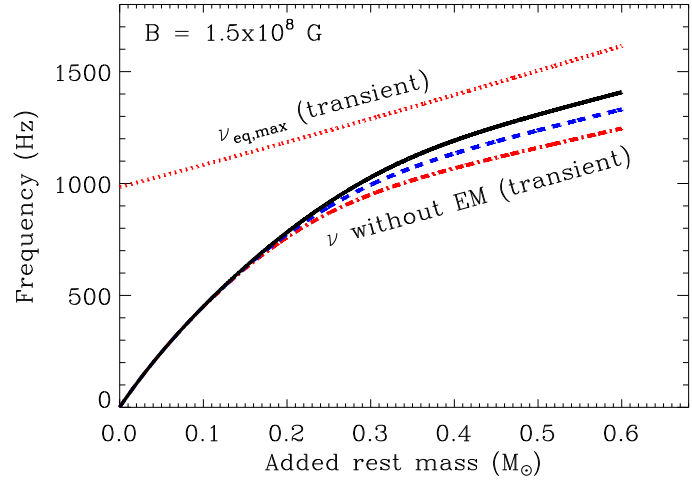}
\caption{Numerically computed evolution of spin frequency as in
	Figure~\ref{fig2}, for different $\eta$ values (see Equation~\ref{Torque8};
	Section~\ref{Torques}). Red dash-dot curve: $\eta = 1$ (as in
	Figure~\ref{fig2}); blue dashed curve: $\eta = 0.5$; black solid
	curve: $\eta = 0.2$. This figure shows that, even for drastically
	different $\eta$ values, the spin frequency evolutions are 
	qualitatively similar to each other, and even quantitatively
	are not very different.
\label{fig4}}
\end{figure}

\begin{figure}[h]
\centering
\includegraphics*[width=12.0cm]{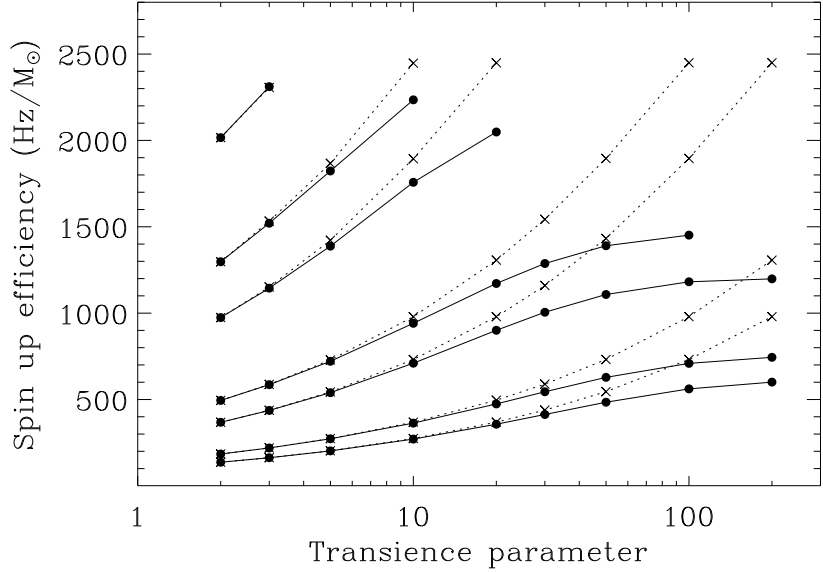}
\caption{Spin-up efficiency (frequency increase per unit
  rest mass transfer) versus transience parameter $\dot{m} =
  \dot{M}_{\rm max}/\dot{M}_{\rm av}$, from numerical computation of
  spin evolution of accreting neutron stars. We assume initial
  parameter values of $M = 1.35 M_\odot$ and $\nu = 1$~Hz and a fixed
  value of $B = 1.5\times10^8$~G.  The solid curves with
  filled circles include the effect of spin-down due to EM torques,
  while the corresponding dotted curves with cross symbols do not.
  For each point (circle or cross), $\Delta M_{\rm tot}=0.6M_\odot$.
  Curves are for the following $\dot{M}_{\rm av}$ 
  values: $5\times10^{14}$, $10^{15}$, $5\times10^{15}$, $10^{16}$,
  $5\times10^{16}$, $10^{17}$, $3\times10^{17}$ g~s$^{-1}$ (bottom to top).
  The $\dot{m}$ values used are $2, 3, 5, 10, 20, 30, 50, 100, 200$
  (left to right).  However, not all of these $\dot{m}$ values are
  available for every $\dot{M}_{\rm av}$ value, since we do not
  consider accretion rates exceeding the Eddington limit. This figure
  shows that spin up efficiency increases with transience for
  reasonable $\dot{M}_{\rm av}$ and $\dot{m}$ ranges, even when
  spin-down due to EM torques is included.
\label{fig5}}
\end{figure}

\begin{figure}[h]
\centering
\includegraphics*[width=8.0cm]{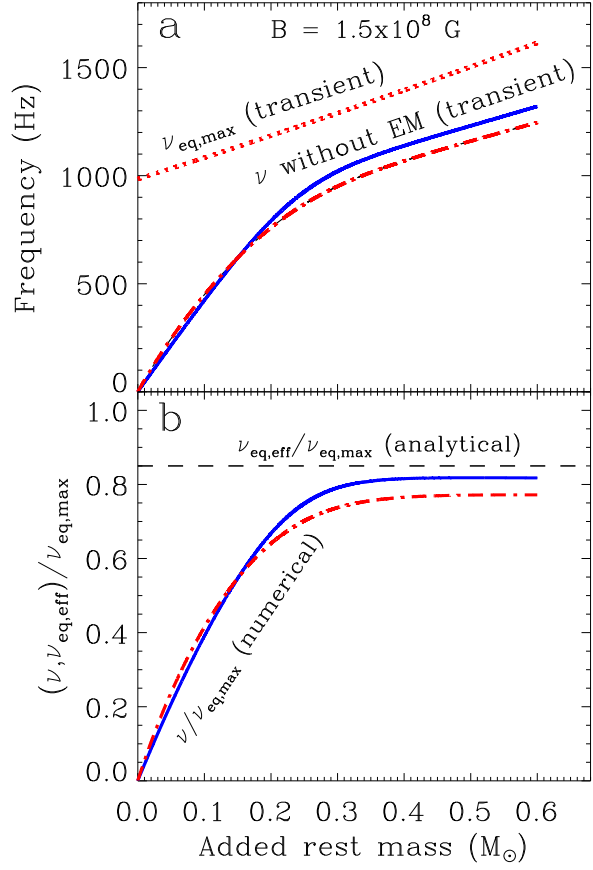}
\caption{
Numerically computed evolution of spin frequency as in Figure 2.
{\it Panel-a}: The red dash-dot curve is for the realistic torques given by 
Equations~\ref{Torque7} and \ref{Torque8}, and the blue solid curve is 
for the approximate torque given by Equation~\ref{Torque12}. Both these
curves show computed spin frequency $\nu$ evolution due to mass transfer.
This panel shows that the spin evolution for the approximate torque is 
qualitatively similar to, and quantitatively only a few percent different from,
that for the realistic torques.
{\it Panel-b}: The red dash-dot and blue solid curves are same as in {\it panel-a},
except the $\nu$ values are normalized with $\nu_{\rm eq,max}$ (shown by the red 
dotted curve in {\it panel-a}). The dashed horizontal line gives a normalized 
(with $\nu_{\rm eq,max}$) analytical value of the equilibrium spin frequency $\nu_{\rm eq, eff}$,
corresponding to Equation~\ref{Eff2}.
This panel shows that both the red dash-dot and blue solid curves saturate
as $\nu$ attains the equilibrium value $\nu_{\rm eq, eff}$. This numerical saturation
value is close to the analytical value of $\approx 0.85$ from
Equation~\ref{Eff2} for $n=6/7$.
\label{fig6}}
\end{figure}

Finally, Figures~\ref{fig2}--\ref{fig6} show that at least some
neutron stars with appropriate parameter values are expected to reach
submillisecond spin rates, even for low $\dot{M}_{\rm av}$.  Our
results reemphasize the puzzling absence of observed MSPs with spin
rates above 716~Hz and suggest a reconsideration of the need for a
competing spin-down mechanism, such as gravitational radiation. Recent
work concluding that the spin equilibrium set by disk-magnetosphere
interaction alone is suffcient to explain the observed spin
distribution \citep[e.g.,][]{Patrunoetal2012b} does not account for
the effect of transient accretion, as we have shown here. Another
recent paper \citet{Parfreyetal2016} proposes a new spin-down
mechanism via an enhanced pulsar wind during the accretion phase, but
again does not consider the effect of transient accretion.  While it
is generally believed that radio pulsar activity only switches on
during an X-ray quiescence phase (Section~\ref{Torques}), these
authors argue that the neutron star magnetic field lines within the
light cylinder can be forced to open to infinity by the accretion
disk, which may give rise to a strong pulsar wind in the accretion
phase. However, even if this possibility is confirmed, spin-down
torques due to gravitational radiation may still play a role for the
fastest rotating MSPs. This is not inconsistent with the absence of
evidence for a gravitational wave torque in slower MSPs
\citep{HaskellPatruno2011}, considering the extremely steep spin
dependence ($\sim \nu^5$; Equation~\ref{GWTorque}) of such torques.
The resulting {\it continuous} gravitational radiation may eventually
itself be directly detectable with interferometric detectors
\citep[e.g.,][]{Aasietal2014}, although the practical obstacles to
making such detections in transient LMXBs (where regular monitoring of
the evolving binary parameters is difficult) have been previously discussed
\citep{Wattsetal2008}.

\acknowledgments
The authors thank an anonymous referee for constructive comments, which were useful to improve
the paper.  D.C. thanks the MIT-India Program for travel support.

\clearpage

\clearpage

\clearpage

\clearpage

\clearpage

\clearpage

\end{document}